\documentclass[aps,prl,floats,twocolumn,amsfonts,amssymb,unsortedaddress,floatfix]{revtex4}
\usepackage{epsfig}
\usepackage{amssymb}
\usepackage{mathtools}

\def\be{\begin{equation}}       \def\ee{\end{equation}}
\def\bea{\begin{eqnarray}}      \def\eea{\end{eqnarray}}

\begin{document}

\title{On the Equivalence of Trapped Colloids, Pinned Vortices, and  Spin Ice} 


\author{Cristiano Nisoli}

\affiliation{Theoretical Division and CNLS, Los Alamos National Laboratory, Los Alamos, NM, 87545, USA}

\begin{abstract}
  We investigate the recently  reported analogies between  pinned vortices in nano-structured superconductors or  colloids in optical traps, and  spin ice materials. The frustration of the two models, one describing colloids and vortices, the other describing spin ice, differs essentially. However,  their effective energetics is made  identical by the contribution of an emergent field associated to a  topological charge. This equivalence extends to the local low-energy  dynamics of the ice manifold, yet  breaks down in lattices of mixed coordination, because of  topological charge transfer between sub-latices.  

\end{abstract}

\maketitle

{\it Introduction.} A  recent multidisciplinary effort in the creation  and study of artificial frustrated nano materials~\cite{Brunner2002, Bohlein2012, Libal2006, Libal2009, Ray2013, Latimer2013, Trastoy2013freezing, Olson2012, Wang2006,  Nisoli2013colloquium, Ke2008,Li2010,  Ke2008a, Nisoli2010, Nisoli2007, Morgan2010, Morgan2013real, Nisoli2012, Ladak2010, Zhang2013,Porro2013, Zhang2011a, Qi2008,Kapaklis2012, Rougemaille2011, Branford2012, Farhan2013, Lammert2010, Morrison2013, Chern2013, Mellado2012, Han2008, Moller2006,Chern2013magnetic}  has  led to the exploration of new of exotic states, including dynamics of magnetic charges and  monopoles~\cite{Castelnovo2008}. 
While artificial spin ice~\cite{Wang2006, Nisoli2013colloquium}, based on magnetic interacting nano structures,  is now a  mature field~\cite{Nisoli2013colloquium},    realizations based on trapped colloids~\cite{Libal2006} and  vortices~\cite{Libal2009, Olson2012} in nano-structured superconductors have been proposed theoretically and realized experimentally~\cite{Latimer2013,Trastoy2013freezing}.  These results show that they  can exhibit a low temperature ice  manifold, seemingly equivalent to the one of spin-ice systems.

In this letter we discuss why and when this equivalence holds.  The  frustration of colloidal systems is of the emergent kind, leading to  an effective energetics of the nodes that includes an emergent field conjugated to the  topological charge. For  spatial modulations,  we show that the equivalence extends to the low energy physics above the ice manifold, mediated by the emergent field. Similarly to spin ice materials, the  ice manifolds are in a Coulomb phase, whereas quasi-ice manifolds are not. This equivalence is based on charge conservation, and it breaks down in lattices of mixed coordination, where a net transfer of topological charge between differently coordinated nodes {\it must} occur---something inherently impossible in magnetic spin ice materials ~\cite{Morrison2013, Chern2013}.

{\it Ice Manifolds.} In water ice each oxygen atom sits at the center of a proton sharing tetrahedron. Two protons are close and covalently bonded, whereas the two others are close to a neighbor:  the ice rule~\cite{Pauling1935}.  
In spin ice materials (natural~\cite{Ramirez1999} or artificial~\cite{Nisoli2013colloquium}) protons are replaced by classical macro spins, and the ice rule   (2 spins pointing in, 2 spins pointing out for a $z=4$ coordination lattice~\cite{Ramirez1999, Wang2006, Nisoli2013colloquium}) or quasi-ice-rule  (1-in/2-out, and 2-in/1-out for $z=3$ lattices~\cite{Qi2008, Nisoli2010, Nisoli2013colloquium}) is dictated by minimization of the  frustrated  energies {\it of the vertices}.  

 \begin{figure}[b!!!]
\begin{center}
\includegraphics[width=.9\columnwidth]{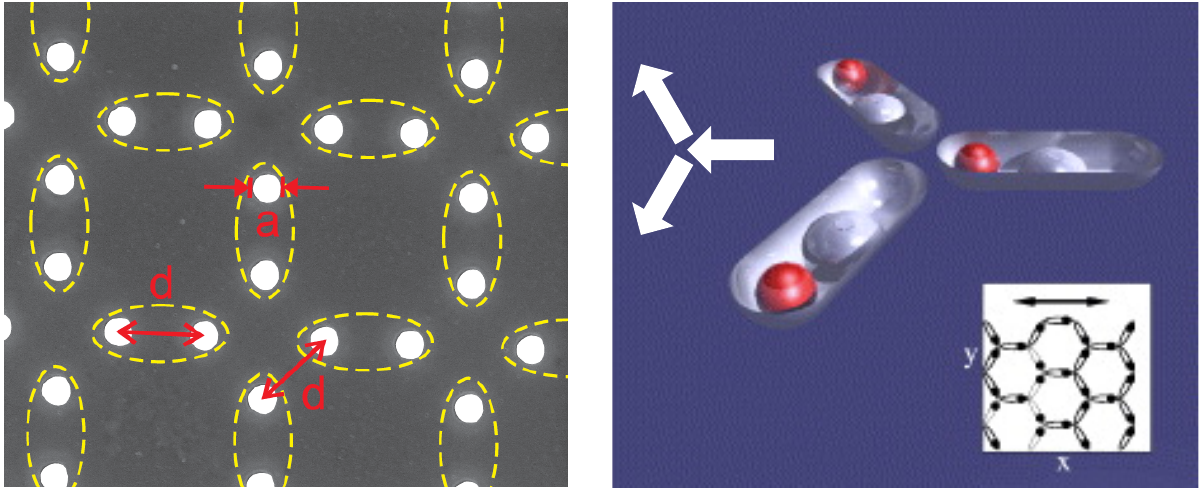}
\caption{Left: SEM image of the nano-patterned substrate for pinning of superconductive vortices in MoGe thin films, from~\cite{Latimer2013}. Right: Schematics of $z=3$, $d=2$ colloidal trap,  corresponding  hexagonal lattice, and spin equivalent, from~\cite{Olson2012}.}
\label{dis}
\end{center}
\end{figure}

This is {\it not} the case for colloids and vortices, for which pairwise interactions in the vertex are instead unfrustrated. Consider  a network of nodes connected by links with coordination $z$. On each link sits a colloid (or a vortex) which only occupy the extreme ends (Fig~\ref{dis}). Each node can have $n=0, 1, \dots z$ close colloids,  and 
\begin{equation}
E_n= E \,\, {n(n-1)}/{2},
\label{en}
\end{equation} 
is the energy of such configurations, each of multiplicity $m_n={z \choose n}$. ($E>0$ as colloids  mutually repel. We neglect interactions---which  are short ranged anyway~\cite{Libal2006}---between different nodes.)

While our system is analogous to a spin ice material with spins directed along the links and pointing toward the colloid (Fig.~1), the energetics in (\ref{en}) 
differ completely. For spin ice,  ice configurations optimize the node energy, and  $n$ is degenerate with $z-n$ because of time reversal symmetry. Indeed in artificial spin ice  the ice rule is accessed even by disjointed vertices~\cite{Li2010}. 
 For colloids, not the pairwise interaction, but rather the allocation of all vertices  in the lowest   
 energy states is frustrated by the lattice 
 hence  an ``emergent'' vertex-frustration. In  spin ice  this only happens in dedicated geometries~\cite{Morrison2013, Chern2013}.

As for spin ice systems we can introduce $\rho_n$, the probability of any node to  be in the  $n$-configuration.  The  ``free energy'' of an uncorrelated gas of nodes  is~\cite{Nisoli2010, Nisoli2007, Morgan2013real}
\begin{equation}
f=\sum_{n=0}^z\left( E_n \rho_n  +T \rho_n \ln \frac{\rho_n}{m_n}\right)-\kappa \left(\sum_{n=0}^z\rho_n-1\right),
\label{f}
\end{equation}
%
 the Lagrange multiplier $T$ representing  an effective~\cite{Nisoli2007,Nisoli2010,Nisoli2012,Morgan2010, Morgan2013real} or real~\cite{Zhang2013, Porro2013} temperature. 
Yet, unlike in spin ice, 
we need to include the   conservation of  colloids in the graph, from which comes  vertex-frustration. The topological charge for a node in the configuration $n$ is
$q_n=2n-z$, 
and is zero for the ice-rule ($n=z/2$). Then, any distribution $\rho_n$ must neutralize the average charge, or  ${\cal Q}=\sum^z_{n=0} q_n \rho_n$=0. Or $ {\cal Q}+{\cal Q}^e=0$ if an extra charge ${\cal Q}^e$ is doped {\it extensively} into the system (see below). 
We thus minimize
\begin{equation}
f_{\mathrm{tot}}=f+\phi\left(\sum_{n=0}^z q_n\rho_n+{\cal Q}^e\right)
\label{fT}
\end{equation}
and obtain, for fixed
 $\phi$, the usual Boltzmann distribution
\begin{equation}
\rho_n={z \choose n}\frac{\exp(- E^{\phi}_n/T)}{Z(T,\phi)},
\label{rho}
\end{equation}
[
$Z(T,\phi)=\sum_n {z \choose n} \exp(- E^{\phi}_n/T)$], in new effective energies $E_n^{\phi}$ which contain an ``electrostatic'' contribution from the emergent field $\phi$ coupled to the  charge $q_n$:
\begin{equation}
E_n^{\phi}=E_n+q_n\phi.
\label{Enfi}
\end{equation}
Optimization of  $f_{\mathrm{tot}}$  with respect to $\phi$ gives
\begin{equation}
{\cal Q}+{\cal Q}^e=T\partial_{\phi}\ln Z(T,\phi)=0,
\label{Q}
\end{equation}
which determines $\phi$ and therefore, through (\ref{Enfi}), $\rho_n$ in (\ref{rho}). 
The temperature-independent choice 
\begin{equation}
\bar \phi/E=-{(z-1)}/{4},
\label{fi}
\end{equation}
 is the solution for a lattice of single coordination and no extensive doping (see below). Indeed it returns  
\begin{equation}
 E_n^{\bar\phi}= E \,[q_n^2+z(z-2)]/8.
\label{Enfibar}
\end{equation}
The last equality in (\ref{Enfibar})   establishes  an ice-like energetics in the absolute value of the topological charges, therefore ensuing ${\cal Q}^e=0$, from $E_n^{\bar\phi}= E_{z-n}^{\bar\phi}$.  Now we can  relabel nodes in terms of   charge rather than colloids, and $\rho_q=\rho_{-q}$.
%
 \begin{figure}[t!!!]
\begin{center}
\includegraphics[width=.9\columnwidth]{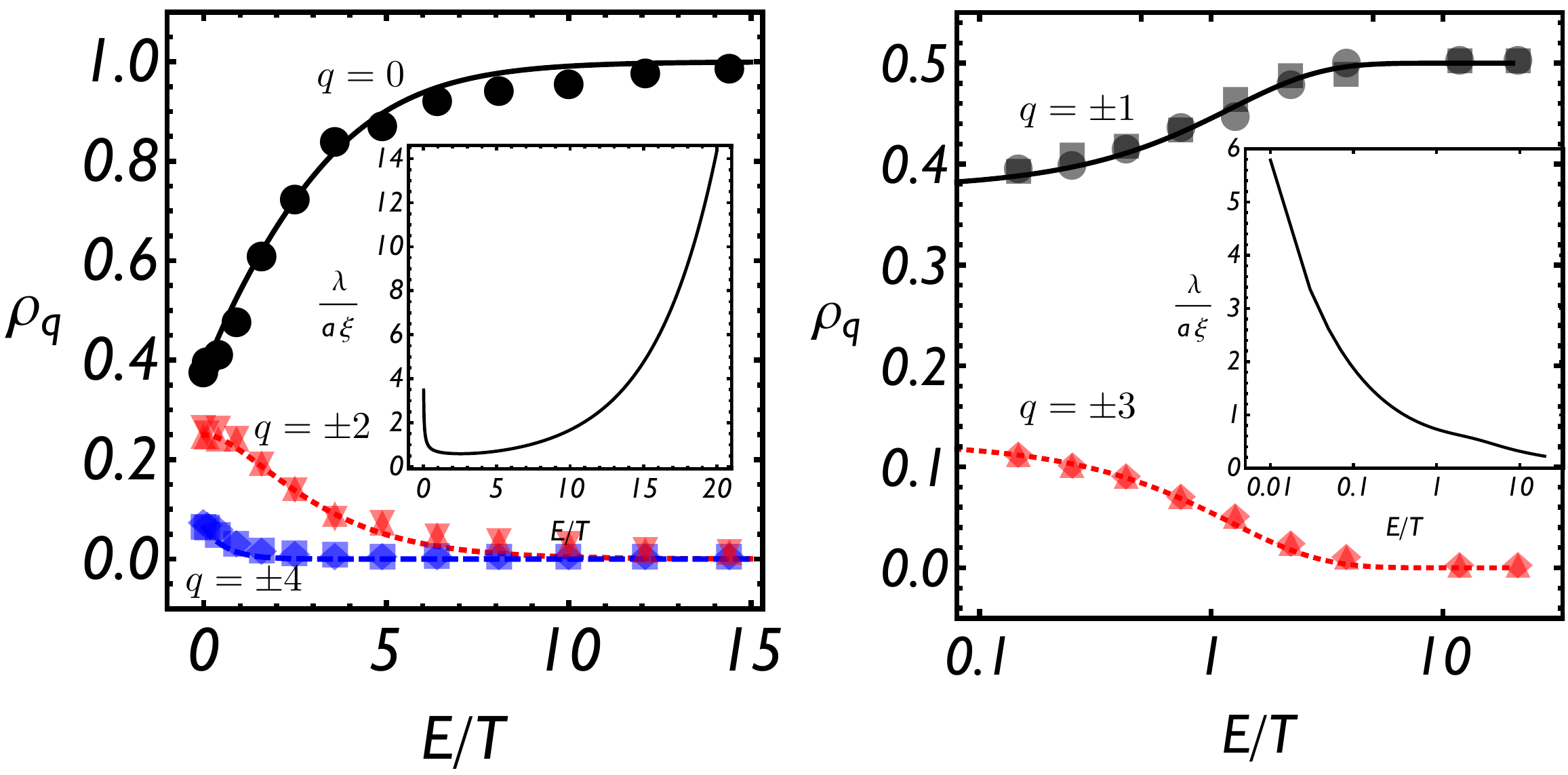}
\caption{Left:  For $z=4$ $\rho_q$ as a function of $E/T$ as in (\ref{rho}) for $q=0$ (black, solid), $q=\pm2$ (red, dotted), $q=\pm4$ (blue, dashed), plotted against numerical data from~\cite{Libal2006} for vertex populations corresponding to $n=2$ ($\bullet$), $n=1$($\blacktriangle$), $n=3$ ($\blacktriangledown$), $n=0$ ($\blacklozenge$), $n=4$ ($\blacksquare$). Right: for $z=3$,  $q=\pm1$ (black, solid), $q=\pm3$ (red, dotted), data from~\cite{Libal2009} for vertex populations corresponding to $n=1$ ($\bullet$), $n=2$($\blacksquare$), $n=3$ ($\blacktriangle$), $n=0$ ($\blacklozenge$). Insets: the screening length as a function of $E/T$ showing the exponential divergence (left)  corresponding to the ice-manifold.}
\label{data}
\end{center}
\end{figure}
%
For  even  $z$, at low temperature the nodes 
enter an ice manifold, with  all vertices tending to the  $n=z/2$ close colloids configuration, or $q=0$ (Fig.~\ref{data}). For odd $z$ a  quasi-ice manifold is approached instead. It  corresponds to embedded  charges $q=\pm 1$ or to $n=(z\pm1)/2$ close colloids, present  in equal proportion. 
We note that the  (quasi)-ice manifold comes from the interaction among colloids: If $E_n$'s were linear in $n$ rather than quadratic in (\ref{en}), they would be subsumed in a redefinition of $\phi$ and vertex frequencies would  follow multiplicity, at any temperature. 

Applications involving 2-D hexagonal lattices of colloids~\cite{Olson2012}  fall into this framework (Fig~\ref{data}). However  the degeneracy of the ice manifold is lifted in practice in other realizations, such as a square lattice of colloids ($z=4,~d=2$)~\cite{Latimer2013,Trastoy2013freezing, Libal2006} which leads to low temperature ordering, as in artificial square ice~\cite{Morgan2010, Zhang2013}. Fig.~\ref{data} shows that  the theory still works, when grouping vertices by topological charge. The further ordering  can then be  accounted for   by symmetry breaking in the energetics in (\ref{en}). Here we concentrate instead on systems that can access a degenerate and thus genuine ice  manifold.  Relevantly, even a square lattice  can be made  degenerate   as proposed by M\"oller~\cite{Moller2006}. 

{\it Extensive Doping.} 
Unlike in spin ice materials, extra charge (colloids) can here be added or subtracted easily; even   mobile charge if the extra colloids or vortices can hop  between links. 
 For pinned vortices in superconductors this corresponds to tweaking the magnetic field around the matching value~\cite{Latimer2013, Trastoy2013freezing}.  
 
If the doping is non-extensive (${\cal Q}^e=0$),  effects are only local (see below), leaving the spatially averaged probabilities unchanged. Conversely, extensively doping    $n^e$  colloids corresponds to an average topological charge per unit vertex ${\cal Q}^e=-2n^e$. This charge breaks the ice, yet not the quasi-ice, manifold.  
If $n^e\ll1$  
we can simply apply  the previous approach with ${\cal Q}^e\ne0$ in (\ref{Q}). Now $\phi$ depends on $T$. From  (\ref{rho}) and (\ref{Enfi}) the behavior of $\rho_n$ at low $T$ is controlled    by the first two terms of the expansion 
\begin{equation}
\phi(T)=\phi_0+\alpha T +O(T^2).
\label{fiexp}
\end{equation}
%
 $\phi_0$ determines the energetics in (\ref{Enfi}), and can be chosen to make either one  or two states  degenerate.  $\alpha$ renormalizes the multiplicities, precisely $m_n \to m_n\exp( q_n\alpha)$, and thus controls the relative admixture of degenerate nodes. 

In lattices of odd $z$, which at low temperature enter a quasi-ice phase of embedded charges $q=\pm1$ in equal proportion,  the extra charge can be screened within the  manifold. Then $\phi_0=-(z-1)E/4$ as in (\ref{fi}) and $\alpha$  fixes the multiplicities of the $q=\pm1$ charges, such that 
$\rho_{q=1} \xrightarrow{T\to 0}(1-{\cal Q}^e)/2, ~~~~\rho_{q=-1} \xrightarrow{T\to 0}(1+{\cal Q}^e)/2$ 
%
and $\rho_{q}\xrightarrow{T\to 0}0$ for $q\ne \pm1$.  
 A genuine ice manifold ($z$ even) however, contains no background charges and extensive excitations   
are needed to absorb the extra charge. Assume ${\cal Q}^e$ positive. We can always chose $\phi_0$ so that the $q=-2$ charges are degenerate with the $q=0$ charges  in the effective  energetics of (\ref{Enfi}). Then $\alpha$ in (\ref{fiexp}) gauges the relative admixture of the two giving
%
$\rho_{q=0}\xrightarrow{T\to 0}1-{\cal Q}^e/2, ~~~~\rho_{q=-2}\xrightarrow{T\to 0} {\cal Q}^e/2$ 
%
and $\rho_{q}\xrightarrow{T\to 0}0$ for $q\ne 0,-2$. The extra charge is screened by  excitations of lowest charge, or $q=-2$. 
This difference in behavior between odd and even coordination number is relevant to lattices of mixed coordination, where a transfer of net topological charge eliminates the ice manifold (below). 

{\it   Charge Screening.} We show now how the emergent field $\phi$  provides information on low energy dynamics and local perturbations. 
If $\rho_n(x)$ is the probability of a node  $x$ to be in configuration $n$, then the  free energy $f$ in (\ref{f}) generalizes to a functional
 \begin{equation}
 {\cal F}[\rho]=\sum_x f(\rho(x))+\Delta {\cal F}[q],
 \label{F}
 \end{equation}
that adds to the uncorrelated local free energy (\ref{f})  the non-local term $\Delta{\cal F}[q]$, which accounts for the effect of the underlying spin structure, including  charge conservation. 
We have already introduced a low temperature approximation since $\Delta{\cal F}$  depends on $\rho_n(x)$ through another (local) functional of $\rho_n(x)$, the density of charge $q(x)=\sum_nq_n\rho_n(x)$.  
To fathom the form of $\Delta {\cal F}[q]$ consider the conjugate field
\begin{equation}
\phi(x)=\frac{\delta \Delta {\cal F}}{\delta q(x)}
\label{jn}
\end{equation}
 and the  Legendre transform
\begin{equation}
{\cal L}[\phi]=\left(\Delta{\cal F}-q \cdot \phi \right)_{q=q[\phi]}
\label{L}
\end{equation}
[where $q \cdot \phi= \sum_{x} q (x)\phi(x)$], which implies
\begin{equation}
q(x)= -\frac{\delta{\cal L}}{\delta \phi(x)}.
\label{qfi}
\end{equation}
and thus finally
\begin{equation}
{\cal F}[\rho,\phi]=\sum_x f(\rho(x))+q \cdot \phi+{\cal L}[\phi].
\label{F2}
\end{equation}
The local functional in (\ref{F2}) looks now more like (\ref{fT}) and the non-local functional ${\cal L}[\phi]$ pertains to the emergent  field, which mediates an entropic interaction. 

We can now construct ${\cal L}$ by perturbing over our previous uncorrelated treatment. Indeed $N_v^{-1}\sum_x \rho_n(x)$ is the probability of {\it any} vertex to have $n$ closed colloids, and should obey ($\ref{rho}$) ($N_v$ is the number of nodes). Then our functionals, restricted to uniform fields, should reduce to the previous treatment. 
From (\ref{jn}),  when $\rho_n(x)$ are uniform, so is $\phi(x)$. Then, in order to recover (\ref{fT}) from (\ref{F2}), 
${\cal L}$,  {\it restricted to  uniform fields}, must be
\begin{equation}
{\cal L}[\phi]=\sum_x{q^e(x)} \phi =N_v {\cal Q}^e \phi
\label{Lunif}
\end{equation}
[$q^e(x)$ is the excess charge in the node $x$ 
and ${\cal Q}^e=N_v^{-1}\sum_x q^e(x)$ is the average excess charge per node]. 

Perturbing over the uniform, average manifold we expand in the derivatives of $\phi$. We assume that  the lattice is regular and allows coarse graining of $x$ into a continuum variable, and thus $\sum_x\to a^{-d}\int_{L^d} d^dx$, where $a^d=L^d/N_v$ is the volume of the unit cell.  At second order 
\begin{equation}
{\cal L}[\phi]=\int_{L^d} \left[q^e \phi -\frac{1}{2}\epsilon \, \partial_i\phi \partial^i \phi\right]\frac{d^dx}{a^d}
\label{L2}
\end{equation}
is the  only admissible form. Indeed,  to be consistent with the uniform solution, second order terms  must be in the derivatives of $\phi$, excluding terms such as $\phi^2$ or $ \partial^i \phi \partial _i q^e$. Here $\epsilon$ is  the generalized permittivity of the emergent field $\phi$ (in general one has  $\epsilon_{ij}$, a suitable  tensor). 

 In taking the functional derivative with respect to $\phi$ in (\ref{qfi}) we cannot discharge derivatives at the boundaries since $\phi$ is not zero at infinity. It is convenient to replace $\phi(x)\to \bar \phi+\phi(x)$, with $\phi \xrightarrow{x\to \infty}0$, and  minimize in both. Minimization in $\phi(x)$ returns
\begin{equation}
-\Delta\phi=(q+ q^e)/\epsilon,
\label{coulomb}
\end{equation}
and thus from (\ref{coulomb}), (\ref{L2}), and (\ref{F2}),
${\cal F}=\int_{L^d} \left[f+ \epsilon \frac{1}{2}\partial_i \phi\partial^i \phi\right]{d^dx}/{a^d}
$ and thus  $\epsilon>0$.  
 Then optimization  of (\ref{F2}) with respect of $\rho(x)$ and $\bar \phi$ leads  again to the charge constraint ${\cal Q} +{\cal Q}^e=0$  for the spatially modulated $\rho_n(x)$ given the Boltzmann law (\ref{rho}) but now with spatially modulated   $\phi (x)$.  
We have now  a Debye-H\"ukel model for an electrolyte solution, where  charges are topological while the interaction $\phi$  is emergent from the   underlying spin network. 

Consider ${\cal Q}^e=0$ but $q^e(x)\ne0$~\footnote{A zero  {\it average extra charge per node} does not imply absence of extra  charge $q^e(x)$, as long as the {\it total charge} $\int_{L^d} q^e(x)d^d x/a^d$ is  non-extensive.}. Then one can define $\rho_n(x)=\rho_n^o+\eta_n(x)$, where $\rho_n^0=\int_{L^d}\rho_n(x)d^dx/a^d$ must be the uniform solution at given $T$ and thus the charge density is $q(x)=\sum_n\eta_n(x) q_n$. Expanding  (\ref{rho}) in $\phi$ around $\bar \phi=-E(z-1)/4$, one finds %
\begin{equation}
\eta_n(x) =- \rho^0_n  q_n  \phi(x)/T
\label{xibar}
\end{equation}
and thus
\begin{equation}
 q(x)=-\overline{{\cal Q}^2} \phi(x)/T
\label{q}
\end{equation}
where $\overline{{\cal Q}^2}(T) =\sum_n \rho^0_n(T)  q_n^2$ is the average  charge fluctuation of the  manifold. Finally from (\ref{coulomb}) and (\ref{q}) $\phi$ satisfies 
\begin{equation}
(\lambda^{-2}-\Delta)\phi=q^e/\epsilon,
\label{phi}
\end{equation}
a screened Poisson equation whose screening length 
\begin{equation}
\lambda=\sqrt{{\epsilon \, T}/{\overline{{\cal Q}^2} }},
\label{lambda}
\end{equation}
 precisely corresponds to the  Debye formula. 

The extra charge is thus locally screened by the fluctuating charges.  But since there is no embedded charge fluctuation in a genuine ice manifold (for even $z$) then $\lambda^{-1}=0$~\footnote{ $\lambda$ approaches infinity exponentially fast in $E/T$ [as $\overline{ {\cal Q}^2} \propto \exp(-E/2T)$] as in Fig~\ref{data}.}, thus revealing an entropic solenoidal (or Coulomb) phase  for $\phi$ in (\ref{phi}). Then standard potential theory in any dimension implies that a mobile extra charge is expelled at the boundaries as the system enters the ice manifold. Conversely for non-zero  $T$, screened point-like charges can diffuse  at an average distance  much larger than $\lambda$.

A (non-extensive) extra charge does not disturb a  genuine ice manifold: indeed from  (\ref{q}), as $\overline{ {\cal Q}^2}=0$,   $q(x)= 0$. 
 Point-like charges   move  in the ice manifold without summoning embedded charges and thus with no alteration to the ice-rule. They do  pair-wise interact via a Laplace Green function. As there is no real transition to an ice manifold, the same applies to charge excitations over the manifold, which interact as the magnetic monopoles~\cite{Castelnovo2008} of spin ice, yet with a difference: because of the short range energetics~\cite{Libal2006}, the interaction between monopoles  is here entirely of the emergent kind and dimensionality dependent: for $d=3$ it is a Coulomb potential and opposite charges are separable, while for $d=2$ they are logarithmically confined. 

Conversely in a quasi-ice manifold (odd $z$) there are always embedded $\pm1$ charges and thus $\overline {{\cal Q}^2}=1$. We have  entropic  screening of  a point-like extra charge  by nearest neighboring embedded charges~\cite{Libal2009}, also seen (numerically) in artificial spin ice of odd coordination~\cite{Chern2013, Chern2013magnetic}. At low temperatures this screening becomes tighter and can form bound states, or polarons~\cite{Chern2013magnetic}. When  spaced at a distance much larger than $\lambda$ they should simply diffuse. 

 \begin{figure}[t!!!]
\begin{center}
\includegraphics[width=.85\columnwidth]{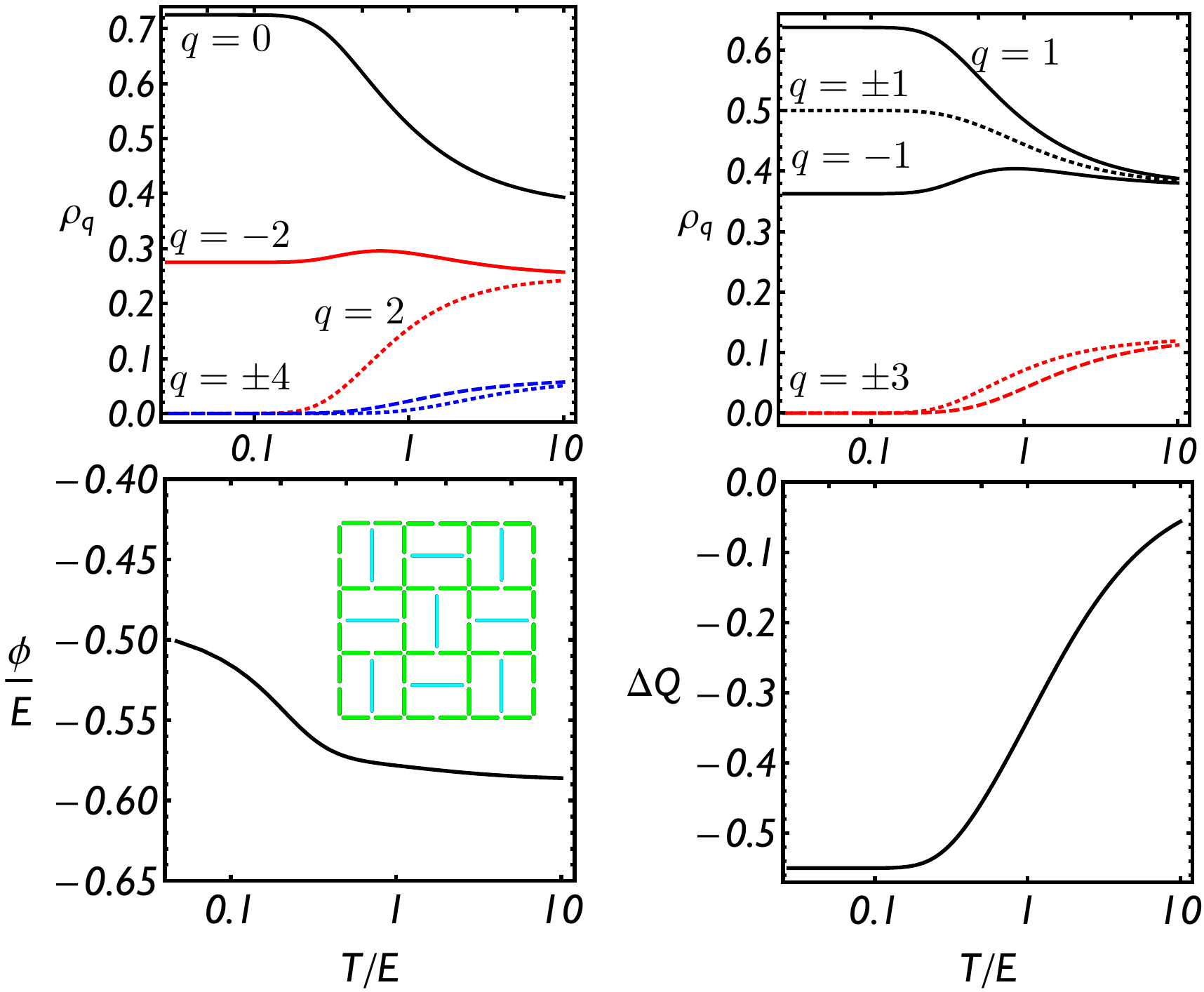}
\caption{For a lattice of mixed coordination there {\it must} be a net transfer of topological charge between vertices of different coordination (bottom right). For $z=3,4$ (inset: the shakti lattice of~\cite{Morrison2013, Chern2013}) $\phi$ at low $T$ approaches its value for $z=3$, corresponding to $E(1-z)/4=-E/2$ (bottom left): thus the sub-vertices of $z=3$  obey the quasi-ice rule (top right, the $q=\pm1$ line, black dotted, is split), but with an imbalance of positive $q=1$ charge, to screen the extra charge coming from the $z=4$ vertices (top left).  }
\label{Mixed}
\end{center}
\end{figure}

As temperature increases, one expects the screening to become less tight. Indeed when $T/E\to \infty$  all links flips independently and vertices are allocated by multiplicity.  From (\ref{rho}) that implies $\phi/T\to 0$, which from (\ref{coulomb}) entails $\epsilon T\to \infty$.  Since $\epsilon$ is inversely proportional to an energy  dimensional considerations fixes it at $\epsilon=\xi^2 a^2 E^{-1}$ where $\xi$ is a number, and is limited in $E/T$. Then (\ref{lambda})  implies, correctly, $\lambda^2/a^2\propto T/E$ (since $\overline{{\cal Q}^2}\le \overline{{\cal Q}^2}|_{T=\infty}=z$). 

%
  
{\it Mixed Coordination and Networks.} When does the equivalence break down? Consider lattices of mixed coordination number, an intriguing scenario that opens a window on more complex geometries~\cite{Morrison2013, Chern2013, Chern2013magnetic} and  in general on dynamics in complex networks, which we will develop elsewhere. Then the free energy 
 is the sum of terms given by (\ref{f}), each corresponding to sub-lattices of different coordination, and weighted by the relative abundance of vertices of that coordination. However, the emergent field must be the same for all sub-lattices: the total charge, not the sub-lattices charge, must be conserved. In fact, there {\it must be}  charge transfer  between sub-lattices of different coordination: since (\ref{fi}) cannot be satisfied for all $z$ simultaneously by the same field,  {\it at most  one} sub-lattice can reach the ice (or quasi-ice) manifold at low $T$, whereas the others  are no longer equivalent to a spin ice system. This is distinctively different from the case of artificial spin ice of mixed coordination which   always enters an ice manifold~\cite{Morrison2013, Chern2013}. There, charge conservation is implied by an  energetics genuinely  degenerate in the sign of the charge.

For definiteness, consider the case of  mixed coordination  4 and 3, in Fig.~\ref{Mixed}.  Our previous discussion on doping  and (\ref{fiexp}) can be employed. If we choose $\phi_0/E =-1/2$ in (\ref{fiexp}), then the $z=3$ sub-lattice  enters the quasi-ice manifold at low $T$. Yet in the $z=4$ sector, from (\ref{Enfi}), $q=-2$ and $q=0$ become degenerate and of lowest effective energy: the $z=4$ vertices  dump positive topological charge on the  $z=3$ ones by  ``exciting'' negative charges ($q=-2$). 

In truth, these are only excitations in the {\it effective} energetics (\ref{Enfi}) for a lattice of single coordination number. For mixed coordination they are are in fact a way to lower the {\it real} energy in (\ref{en}), whenever geometry permits. Then the $z=3$ sub-lattice  can screen the extra charge without abandoning the quasi-ice manifold, in a way reminiscent of
 what happens to the shakti~\cite{Morrison2013, Chern2013, Gilbert2014}  and pentagonal~\cite{Chern2013magnetic} artificial spin ice {\it above} their ice manifold. 
  
 
{\it Conclusion.} We have studied the equivalence between systems of trapped colloids or pinned vortices,  and  spin ice: their ensemble is controlled by an effective energetics that accounts for the effect of an emergent field on the topological charges of the vertices. We find that lattices of even coordination number can access an ice manifold, which is a Coulomb phase that for $d\ge 3$ can support separable monopolar excitations, as in  spin ice. Lattices of odd coordination  access a quasi-ice manifold, as also seen in artificial spin ice, in which polarons can form. Finally the equivalence breaks down in lattices of mixed coordination, whose behavior is essentially different from mixed coordination spin ices. 

We thank P. E. Lammert for  reading the manuscript, C. \& C. Reichhardt and A. Libal for providing  numerical data. This work was carried out under the auspices of the National Nuclear Security Administration of the U.S. Department of Energy at Los Alamos National Laboratory under Contract No.~DEAC52-06NA25396.

\bibliography{library.bib}{}

\begin{thebibliography}{38}
\expandafter\ifx\csname natexlab\endcsname\relax\def\natexlab#1{#1}\fi
\expandafter\ifx\csname bibnamefont\endcsname\relax
  \def\bibnamefont#1{#1}\fi
\expandafter\ifx\csname bibfnamefont\endcsname\relax
  \def\bibfnamefont#1{#1}\fi
\expandafter\ifx\csname citenamefont\endcsname\relax
  \def\citenamefont#1{#1}\fi
\expandafter\ifx\csname url\endcsname\relax
  \def\url#1{\texttt{#1}}\fi
\expandafter\ifx\csname urlprefix\endcsname\relax\def\urlprefix{URL }\fi
\providecommand{\bibinfo}[2]{#2}
\providecommand{\eprint}[2][]{\url{#2}}

\bibitem[{\citenamefont{Brunner and Bechinger}(2002)}]{Brunner2002}
\bibinfo{author}{\bibfnamefont{M.}~\bibnamefont{Brunner}} \bibnamefont{and}
  \bibinfo{author}{\bibfnamefont{C.}~\bibnamefont{Bechinger}},
  \bibinfo{journal}{Physical Review Letters} \textbf{\bibinfo{volume}{88}},
  \bibinfo{pages}{248302} (\bibinfo{year}{2002}).

\bibitem[{\citenamefont{Bohlein et~al.}(2012)\citenamefont{Bohlein, Mikhael,
  and Bechinger}}]{Bohlein2012}
\bibinfo{author}{\bibfnamefont{T.}~\bibnamefont{Bohlein}},
  \bibinfo{author}{\bibfnamefont{J.}~\bibnamefont{Mikhael}}, \bibnamefont{and}
  \bibinfo{author}{\bibfnamefont{C.}~\bibnamefont{Bechinger}},
  \bibinfo{journal}{Nature materials} \textbf{\bibinfo{volume}{11}},
  \bibinfo{pages}{126} (\bibinfo{year}{2012}).

\bibitem[{\citenamefont{Lib\'{a}l et~al.}(2006)\citenamefont{Lib\'{a}l,
  Reichhardt, and {Olson Reichhardt}}}]{Libal2006}
\bibinfo{author}{\bibfnamefont{A.}~\bibnamefont{Lib\'{a}l}},
  \bibinfo{author}{\bibfnamefont{C.}~\bibnamefont{Reichhardt}},
  \bibnamefont{and} \bibinfo{author}{\bibfnamefont{C.~J.} \bibnamefont{{Olson
  Reichhardt}}}, \bibinfo{journal}{Phys. Rev. Lett.}
  \textbf{\bibinfo{volume}{97}}, \bibinfo{pages}{228302}
  (\bibinfo{year}{2006}).

\bibitem[{\citenamefont{Lib{\'a}l et~al.}(2009)\citenamefont{Lib{\'a}l,
  Reichhardt, and Reichhardt}}]{Libal2009}
\bibinfo{author}{\bibfnamefont{A.}~\bibnamefont{Lib{\'a}l}},
  \bibinfo{author}{\bibfnamefont{C.~O.} \bibnamefont{Reichhardt}},
  \bibnamefont{and}
  \bibinfo{author}{\bibfnamefont{C.}~\bibnamefont{Reichhardt}},
  \bibinfo{journal}{Physical review letters} \textbf{\bibinfo{volume}{102}},
  \bibinfo{pages}{237004} (\bibinfo{year}{2009}).

\bibitem[{\citenamefont{Ray et~al.}(2013)\citenamefont{Ray, Reichhardt,
  Jank{\'o}, and Reichhardt}}]{Ray2013}
\bibinfo{author}{\bibfnamefont{D.}~\bibnamefont{Ray}},
  \bibinfo{author}{\bibfnamefont{C.~O.} \bibnamefont{Reichhardt}},
  \bibinfo{author}{\bibfnamefont{B.}~\bibnamefont{Jank{\'o}}},
  \bibnamefont{and}
  \bibinfo{author}{\bibfnamefont{C.}~\bibnamefont{Reichhardt}},
  \bibinfo{journal}{Physical review letters} \textbf{\bibinfo{volume}{110}},
  \bibinfo{pages}{267001} (\bibinfo{year}{2013}).

\bibitem[{\citenamefont{Latimer et~al.}(2013)\citenamefont{Latimer, Berdiyorov,
  Xiao, Peeters, and Kwok}}]{Latimer2013}
\bibinfo{author}{\bibfnamefont{M.~L.} \bibnamefont{Latimer}},
  \bibinfo{author}{\bibfnamefont{G.~R.} \bibnamefont{Berdiyorov}},
  \bibinfo{author}{\bibfnamefont{Z.~L.} \bibnamefont{Xiao}},
  \bibinfo{author}{\bibfnamefont{F.~M.} \bibnamefont{Peeters}},
  \bibnamefont{and} \bibinfo{author}{\bibfnamefont{W.~K.} \bibnamefont{Kwok}},
  \bibinfo{journal}{Phys. Rev. Lett.} \textbf{\bibinfo{volume}{111}},
  \bibinfo{pages}{067001} (\bibinfo{year}{2013}).

\bibitem[{\citenamefont{Trastoy et~al.}(2013)\citenamefont{Trastoy, Malnou,
  Ulysse, Bernard, Bergeal, Faini, Lesueur, Briatico, and
  Villegas}}]{Trastoy2013freezing}
\bibinfo{author}{\bibfnamefont{J.}~\bibnamefont{Trastoy}},
  \bibinfo{author}{\bibfnamefont{M.}~\bibnamefont{Malnou}},
  \bibinfo{author}{\bibfnamefont{C.}~\bibnamefont{Ulysse}},
  \bibinfo{author}{\bibfnamefont{R.}~\bibnamefont{Bernard}},
  \bibinfo{author}{\bibfnamefont{N.}~\bibnamefont{Bergeal}},
  \bibinfo{author}{\bibfnamefont{G.}~\bibnamefont{Faini}},
  \bibinfo{author}{\bibfnamefont{J.}~\bibnamefont{Lesueur}},
  \bibinfo{author}{\bibfnamefont{J.}~\bibnamefont{Briatico}}, \bibnamefont{and}
  \bibinfo{author}{\bibfnamefont{J.~E.} \bibnamefont{Villegas}},
  \bibinfo{journal}{arXiv preprint arXiv:1307.2881}  (\bibinfo{year}{2013}).

\bibitem[{\citenamefont{Reichhardt et~al.}(2012)\citenamefont{Reichhardt,
  Libál, and Reichhardt}}]{Olson2012}
\bibinfo{author}{\bibfnamefont{C.~J.~O.} \bibnamefont{Reichhardt}},
  \bibinfo{author}{\bibfnamefont{A.}~\bibnamefont{Libál}}, \bibnamefont{and}
  \bibinfo{author}{\bibfnamefont{C.}~\bibnamefont{Reichhardt}},
  \bibinfo{journal}{New Journal of Physics} \textbf{\bibinfo{volume}{14}},
  \bibinfo{pages}{025006} (\bibinfo{year}{2012}).

\bibitem[{\citenamefont{Wang et~al.}(2006)\citenamefont{Wang, Nisoli, Freitas,
  Li, McConville, Cooley, Lund, Samarth, Leighton, Crespi et~al.}}]{Wang2006}
\bibinfo{author}{\bibfnamefont{R.~F.} \bibnamefont{Wang}},
  \bibinfo{author}{\bibfnamefont{C.}~\bibnamefont{Nisoli}},
  \bibinfo{author}{\bibfnamefont{R.~S.} \bibnamefont{Freitas}},
  \bibinfo{author}{\bibfnamefont{J.}~\bibnamefont{Li}},
  \bibinfo{author}{\bibfnamefont{W.}~\bibnamefont{McConville}},
  \bibinfo{author}{\bibfnamefont{B.~J.} \bibnamefont{Cooley}},
  \bibinfo{author}{\bibfnamefont{M.~S.} \bibnamefont{Lund}},
  \bibinfo{author}{\bibfnamefont{N.}~\bibnamefont{Samarth}},
  \bibinfo{author}{\bibfnamefont{C.}~\bibnamefont{Leighton}},
  \bibinfo{author}{\bibfnamefont{V.~H.} \bibnamefont{Crespi}},
  \bibnamefont{et~al.}, \bibinfo{journal}{Nature}
  \textbf{\bibinfo{volume}{439}}, \bibinfo{pages}{303} (\bibinfo{year}{2006}).

\bibitem[{\citenamefont{Nisoli et~al.}(2013)\citenamefont{Nisoli, Moessner, and
  Schiffer}}]{Nisoli2013colloquium}
\bibinfo{author}{\bibfnamefont{C.}~\bibnamefont{Nisoli}},
  \bibinfo{author}{\bibfnamefont{R.}~\bibnamefont{Moessner}}, \bibnamefont{and}
  \bibinfo{author}{\bibfnamefont{P.}~\bibnamefont{Schiffer}},
  \bibinfo{journal}{Reviews of Modern Physics} \textbf{\bibinfo{volume}{85}},
  \bibinfo{pages}{1473} (\bibinfo{year}{2013}).

\bibitem[{\citenamefont{Ke et~al.}(2008{\natexlab{a}})\citenamefont{Ke, Li,
  Nisoli, Lammert, McConville, Wang, Crespi, and Schiffer}}]{Ke2008}
\bibinfo{author}{\bibfnamefont{X.}~\bibnamefont{Ke}},
  \bibinfo{author}{\bibfnamefont{J.}~\bibnamefont{Li}},
  \bibinfo{author}{\bibfnamefont{C.}~\bibnamefont{Nisoli}},
  \bibinfo{author}{\bibfnamefont{P.~E.} \bibnamefont{Lammert}},
  \bibinfo{author}{\bibfnamefont{W.}~\bibnamefont{McConville}},
  \bibinfo{author}{\bibfnamefont{R.}~\bibnamefont{Wang}},
  \bibinfo{author}{\bibfnamefont{V.~H.} \bibnamefont{Crespi}},
  \bibnamefont{and} \bibinfo{author}{\bibfnamefont{P.}~\bibnamefont{Schiffer}},
  \bibinfo{journal}{Phys. Rev. Lett.} \textbf{\bibinfo{volume}{101}},
  \bibinfo{pages}{037205} (\bibinfo{year}{2008}{\natexlab{a}}).

\bibitem[{\citenamefont{Li et~al.}(2010)\citenamefont{Li, Zhang, Bartell,
  Nisoli, Ke, Lammert, Crespi, and Schiffer}}]{Li2010}
\bibinfo{author}{\bibfnamefont{J.}~\bibnamefont{Li}},
  \bibinfo{author}{\bibfnamefont{S.}~\bibnamefont{Zhang}},
  \bibinfo{author}{\bibfnamefont{J.}~\bibnamefont{Bartell}},
  \bibinfo{author}{\bibfnamefont{C.}~\bibnamefont{Nisoli}},
  \bibinfo{author}{\bibfnamefont{X.}~\bibnamefont{Ke}},
  \bibinfo{author}{\bibfnamefont{P.}~\bibnamefont{Lammert}},
  \bibinfo{author}{\bibfnamefont{V.}~\bibnamefont{Crespi}}, \bibnamefont{and}
  \bibinfo{author}{\bibfnamefont{P.}~\bibnamefont{Schiffer}},
  \bibinfo{journal}{Phys. Rev. B} \textbf{\bibinfo{volume}{82}},
  \bibinfo{pages}{134407} (\bibinfo{year}{2010}).

\bibitem[{\citenamefont{Ke et~al.}(2008{\natexlab{b}})\citenamefont{Ke, Li,
  Zhang, Nisoli, Crespi, and Schiffer}}]{Ke2008a}
\bibinfo{author}{\bibfnamefont{X.}~\bibnamefont{Ke}},
  \bibinfo{author}{\bibfnamefont{J.}~\bibnamefont{Li}},
  \bibinfo{author}{\bibfnamefont{S.}~\bibnamefont{Zhang}},
  \bibinfo{author}{\bibfnamefont{C.}~\bibnamefont{Nisoli}},
  \bibinfo{author}{\bibfnamefont{V.~H.} \bibnamefont{Crespi}},
  \bibnamefont{and} \bibinfo{author}{\bibfnamefont{P.}~\bibnamefont{Schiffer}},
  \bibinfo{journal}{Appl. Phys. Lett.} \textbf{\bibinfo{volume}{93}},
  \bibinfo{pages}{252504} (\bibinfo{year}{2008}{\natexlab{b}}).

\bibitem[{\citenamefont{Nisoli et~al.}(2010)\citenamefont{Nisoli, Li, Ke,
  Garand, Schiffer, and Crespi}}]{Nisoli2010}
\bibinfo{author}{\bibfnamefont{C.}~\bibnamefont{Nisoli}},
  \bibinfo{author}{\bibfnamefont{J.}~\bibnamefont{Li}},
  \bibinfo{author}{\bibfnamefont{X.}~\bibnamefont{Ke}},
  \bibinfo{author}{\bibfnamefont{D.}~\bibnamefont{Garand}},
  \bibinfo{author}{\bibfnamefont{P.}~\bibnamefont{Schiffer}}, \bibnamefont{and}
  \bibinfo{author}{\bibfnamefont{V.~H.} \bibnamefont{Crespi}},
  \bibinfo{journal}{Phys. Rev. Lett.} \textbf{\bibinfo{volume}{105}},
  \bibinfo{pages}{047205} (\bibinfo{year}{2010}).

\bibitem[{\citenamefont{Nisoli et~al.}(2007)\citenamefont{Nisoli, Wang, Li,
  McConville, Lammert, Schiffer, and Crespi}}]{Nisoli2007}
\bibinfo{author}{\bibfnamefont{C.}~\bibnamefont{Nisoli}},
  \bibinfo{author}{\bibfnamefont{R.}~\bibnamefont{Wang}},
  \bibinfo{author}{\bibfnamefont{J.}~\bibnamefont{Li}},
  \bibinfo{author}{\bibfnamefont{W.}~\bibnamefont{McConville}},
  \bibinfo{author}{\bibfnamefont{P.}~\bibnamefont{Lammert}},
  \bibinfo{author}{\bibfnamefont{P.}~\bibnamefont{Schiffer}}, \bibnamefont{and}
  \bibinfo{author}{\bibfnamefont{V.}~\bibnamefont{Crespi}},
  \bibinfo{journal}{Phys. Rev. Lett.} \textbf{\bibinfo{volume}{98}},
  \bibinfo{pages}{217203} (\bibinfo{year}{2007}).

\bibitem[{\citenamefont{Morgan et~al.}(2010)\citenamefont{Morgan, Stein,
  Langridge, and Marrows}}]{Morgan2010}
\bibinfo{author}{\bibfnamefont{J.~P.} \bibnamefont{Morgan}},
  \bibinfo{author}{\bibfnamefont{A.}~\bibnamefont{Stein}},
  \bibinfo{author}{\bibfnamefont{S.}~\bibnamefont{Langridge}},
  \bibnamefont{and} \bibinfo{author}{\bibfnamefont{C.~H.}
  \bibnamefont{Marrows}}, \bibinfo{journal}{Nat. Phys.}
  \textbf{\bibinfo{volume}{7}}, \bibinfo{pages}{75} (\bibinfo{year}{2010}).

\bibitem[{\citenamefont{Morgan et~al.}(2013)\citenamefont{Morgan, Akerman,
  Stein, Phatak, Evans, Langridge, and Marrows}}]{Morgan2013real}
\bibinfo{author}{\bibfnamefont{J.~P.} \bibnamefont{Morgan}},
  \bibinfo{author}{\bibfnamefont{J.}~\bibnamefont{Akerman}},
  \bibinfo{author}{\bibfnamefont{A.}~\bibnamefont{Stein}},
  \bibinfo{author}{\bibfnamefont{C.}~\bibnamefont{Phatak}},
  \bibinfo{author}{\bibfnamefont{R.}~\bibnamefont{Evans}},
  \bibinfo{author}{\bibfnamefont{S.}~\bibnamefont{Langridge}},
  \bibnamefont{and} \bibinfo{author}{\bibfnamefont{C.~H.}
  \bibnamefont{Marrows}}, \bibinfo{journal}{Physical Review B}
  \textbf{\bibinfo{volume}{87}}, \bibinfo{pages}{024405}
  (\bibinfo{year}{2013}).

\bibitem[{\citenamefont{Nisoli}(2012)}]{Nisoli2012}
\bibinfo{author}{\bibfnamefont{C.}~\bibnamefont{Nisoli}}, \bibinfo{journal}{New
  J. Phys.} \textbf{\bibinfo{volume}{14}}, \bibinfo{pages}{035017}
  (\bibinfo{year}{2012}).

\bibitem[{\citenamefont{{Ladak} et~al.}(2010)\citenamefont{{Ladak}, {Read},
  {Perkins}, {Cohen}, and {Branford}}}]{Ladak2010}
\bibinfo{author}{\bibfnamefont{S.}~\bibnamefont{{Ladak}}},
  \bibinfo{author}{\bibfnamefont{D.~E.} \bibnamefont{{Read}}},
  \bibinfo{author}{\bibfnamefont{G.~K.} \bibnamefont{{Perkins}}},
  \bibinfo{author}{\bibfnamefont{L.~F.} \bibnamefont{{Cohen}}},
  \bibnamefont{and} \bibinfo{author}{\bibfnamefont{W.~R.}
  \bibnamefont{{Branford}}}, \bibinfo{journal}{Nature Physics}
  \textbf{\bibinfo{volume}{6}}, \bibinfo{pages}{359} (\bibinfo{year}{2010}).

\bibitem[{\citenamefont{Zhang et~al.}(2013)\citenamefont{Zhang, Gilbert,
  Nisoli, Chern, Erickson, O’Brien, Leighton, Lammert, Crespi, and
  Schiffer}}]{Zhang2013}
\bibinfo{author}{\bibfnamefont{S.}~\bibnamefont{Zhang}},
  \bibinfo{author}{\bibfnamefont{I.}~\bibnamefont{Gilbert}},
  \bibinfo{author}{\bibfnamefont{C.}~\bibnamefont{Nisoli}},
  \bibinfo{author}{\bibfnamefont{G.-W.} \bibnamefont{Chern}},
  \bibinfo{author}{\bibfnamefont{M.~J.} \bibnamefont{Erickson}},
  \bibinfo{author}{\bibfnamefont{L.}~\bibnamefont{O’Brien}},
  \bibinfo{author}{\bibfnamefont{C.}~\bibnamefont{Leighton}},
  \bibinfo{author}{\bibfnamefont{P.~E.} \bibnamefont{Lammert}},
  \bibinfo{author}{\bibfnamefont{V.~H.} \bibnamefont{Crespi}},
  \bibnamefont{and} \bibinfo{author}{\bibfnamefont{P.}~\bibnamefont{Schiffer}},
  \bibinfo{journal}{Nature} \textbf{\bibinfo{volume}{500}},
  \bibinfo{pages}{553} (\bibinfo{year}{2013}).

\bibitem[{\citenamefont{Porro et~al.}(2013)\citenamefont{Porro, Bedoya-Pinto,
  Berger, and Vavassori}}]{Porro2013}
\bibinfo{author}{\bibfnamefont{J.}~\bibnamefont{Porro}},
  \bibinfo{author}{\bibfnamefont{A.}~\bibnamefont{Bedoya-Pinto}},
  \bibinfo{author}{\bibfnamefont{A.}~\bibnamefont{Berger}}, \bibnamefont{and}
  \bibinfo{author}{\bibfnamefont{P.}~\bibnamefont{Vavassori}},
  \bibinfo{journal}{New Journal of Physics} \textbf{\bibinfo{volume}{15}},
  \bibinfo{pages}{055012} (\bibinfo{year}{2013}).

\bibitem[{\citenamefont{Zhang et~al.}(2011)\citenamefont{Zhang, Li, Bartell,
  Ke, Nisoli, Lammert, Crespi, and Schiffer}}]{Zhang2011a}
\bibinfo{author}{\bibfnamefont{S.}~\bibnamefont{Zhang}},
  \bibinfo{author}{\bibfnamefont{J.}~\bibnamefont{Li}},
  \bibinfo{author}{\bibfnamefont{J.}~\bibnamefont{Bartell}},
  \bibinfo{author}{\bibfnamefont{X.}~\bibnamefont{Ke}},
  \bibinfo{author}{\bibfnamefont{C.}~\bibnamefont{Nisoli}},
  \bibinfo{author}{\bibfnamefont{P.}~\bibnamefont{Lammert}},
  \bibinfo{author}{\bibfnamefont{V.}~\bibnamefont{Crespi}}, \bibnamefont{and}
  \bibinfo{author}{\bibfnamefont{P.}~\bibnamefont{Schiffer}},
  \bibinfo{journal}{Phys. Rev. Lett.} \textbf{\bibinfo{volume}{107}},
  \bibinfo{pages}{117204} (\bibinfo{year}{2011}).

\bibitem[{\citenamefont{Qi et~al.}(2008)\citenamefont{Qi, Brintlinger, and
  Cumings}}]{Qi2008}
\bibinfo{author}{\bibfnamefont{Y.}~\bibnamefont{Qi}},
  \bibinfo{author}{\bibfnamefont{T.}~\bibnamefont{Brintlinger}},
  \bibnamefont{and} \bibinfo{author}{\bibfnamefont{J.}~\bibnamefont{Cumings}},
  \bibinfo{journal}{Phys. Rev. B} \textbf{\bibinfo{volume}{77}},
  \bibinfo{pages}{094418} (\bibinfo{year}{2008}).

\bibitem[{\citenamefont{Kapaklis et~al.}(2012)\citenamefont{Kapaklis, Arnalds,
  Harman-Clarke, Papaioannou, Karimipour, Korelis, Taroni, Holdsworth,
  Bramwell, and Hj\"{o}rvarsson}}]{Kapaklis2012}
\bibinfo{author}{\bibfnamefont{V.}~\bibnamefont{Kapaklis}},
  \bibinfo{author}{\bibfnamefont{U.~B.} \bibnamefont{Arnalds}},
  \bibinfo{author}{\bibfnamefont{A.}~\bibnamefont{Harman-Clarke}},
  \bibinfo{author}{\bibfnamefont{E.~T.} \bibnamefont{Papaioannou}},
  \bibinfo{author}{\bibfnamefont{M.}~\bibnamefont{Karimipour}},
  \bibinfo{author}{\bibfnamefont{P.}~\bibnamefont{Korelis}},
  \bibinfo{author}{\bibfnamefont{A.}~\bibnamefont{Taroni}},
  \bibinfo{author}{\bibfnamefont{P.~C.~W.} \bibnamefont{Holdsworth}},
  \bibinfo{author}{\bibfnamefont{S.~T.} \bibnamefont{Bramwell}},
  \bibnamefont{and}
  \bibinfo{author}{\bibfnamefont{B.}~\bibnamefont{Hj\"{o}rvarsson}},
  \bibinfo{journal}{New J. Phys.} \textbf{\bibinfo{volume}{14}},
  \bibinfo{pages}{035009} (\bibinfo{year}{2012}).

\bibitem[{\citenamefont{Rougemaille et~al.}(2011)\citenamefont{Rougemaille,
  Montaigne, Canals, Duluard, Lacour, Hehn, Belkhou, Fruchart, El~Moussaoui,
  Bendounan et~al.}}]{Rougemaille2011}
\bibinfo{author}{\bibfnamefont{N.}~\bibnamefont{Rougemaille}},
  \bibinfo{author}{\bibfnamefont{F.}~\bibnamefont{Montaigne}},
  \bibinfo{author}{\bibfnamefont{B.}~\bibnamefont{Canals}},
  \bibinfo{author}{\bibfnamefont{A.}~\bibnamefont{Duluard}},
  \bibinfo{author}{\bibfnamefont{D.}~\bibnamefont{Lacour}},
  \bibinfo{author}{\bibfnamefont{M.}~\bibnamefont{Hehn}},
  \bibinfo{author}{\bibfnamefont{R.}~\bibnamefont{Belkhou}},
  \bibinfo{author}{\bibfnamefont{O.}~\bibnamefont{Fruchart}},
  \bibinfo{author}{\bibfnamefont{S.}~\bibnamefont{El~Moussaoui}},
  \bibinfo{author}{\bibfnamefont{A.}~\bibnamefont{Bendounan}},
  \bibnamefont{et~al.}, \bibinfo{journal}{Phys. Rev. Lett.}
  \textbf{\bibinfo{volume}{106}}, \bibinfo{pages}{057209}
  (\bibinfo{year}{2011}).

\bibitem[{\citenamefont{Branford et~al.}(2012)\citenamefont{Branford, Ladak,
  Read, Zeissler, and Cohen}}]{Branford2012}
\bibinfo{author}{\bibfnamefont{W.~R.} \bibnamefont{Branford}},
  \bibinfo{author}{\bibfnamefont{S.}~\bibnamefont{Ladak}},
  \bibinfo{author}{\bibfnamefont{D.~E.} \bibnamefont{Read}},
  \bibinfo{author}{\bibfnamefont{K.}~\bibnamefont{Zeissler}}, \bibnamefont{and}
  \bibinfo{author}{\bibfnamefont{L.~F.} \bibnamefont{Cohen}},
  \bibinfo{journal}{Science} \textbf{\bibinfo{volume}{335}},
  \bibinfo{pages}{1597} (\bibinfo{year}{2012}).

\bibitem[{\citenamefont{Farhan et~al.}(2013)\citenamefont{Farhan, Derlet,
  Kleibert, Balan, Chopdekar, Wyss, Anghinolfi, Nolting, and
  Heyderman}}]{Farhan2013}
\bibinfo{author}{\bibfnamefont{A.}~\bibnamefont{Farhan}},
  \bibinfo{author}{\bibfnamefont{P.}~\bibnamefont{Derlet}},
  \bibinfo{author}{\bibfnamefont{A.}~\bibnamefont{Kleibert}},
  \bibinfo{author}{\bibfnamefont{A.}~\bibnamefont{Balan}},
  \bibinfo{author}{\bibfnamefont{R.}~\bibnamefont{Chopdekar}},
  \bibinfo{author}{\bibfnamefont{M.}~\bibnamefont{Wyss}},
  \bibinfo{author}{\bibfnamefont{L.}~\bibnamefont{Anghinolfi}},
  \bibinfo{author}{\bibfnamefont{F.}~\bibnamefont{Nolting}}, \bibnamefont{and}
  \bibinfo{author}{\bibfnamefont{L.}~\bibnamefont{Heyderman}},
  \bibinfo{journal}{Nature Physics}  (\bibinfo{year}{2013}).

\bibitem[{\citenamefont{Lammert et~al.}(2010)\citenamefont{Lammert, Ke, Li,
  Nisoli, Garand, Crespi, and Schiffer}}]{Lammert2010}
\bibinfo{author}{\bibfnamefont{P.~E.} \bibnamefont{Lammert}},
  \bibinfo{author}{\bibfnamefont{X.}~\bibnamefont{Ke}},
  \bibinfo{author}{\bibfnamefont{J.}~\bibnamefont{Li}},
  \bibinfo{author}{\bibfnamefont{C.}~\bibnamefont{Nisoli}},
  \bibinfo{author}{\bibfnamefont{D.~M.} \bibnamefont{Garand}},
  \bibinfo{author}{\bibfnamefont{V.~H.} \bibnamefont{Crespi}},
  \bibnamefont{and} \bibinfo{author}{\bibfnamefont{P.}~\bibnamefont{Schiffer}},
  \bibinfo{journal}{Nat. Phys.} \textbf{\bibinfo{volume}{6}},
  \bibinfo{pages}{786} (\bibinfo{year}{2010}).

\bibitem[{\citenamefont{Morrison et~al.}(2013)\citenamefont{Morrison, Nelson,
  and Nisoli}}]{Morrison2013}
\bibinfo{author}{\bibfnamefont{M.~J.} \bibnamefont{Morrison}},
  \bibinfo{author}{\bibfnamefont{T.~R.} \bibnamefont{Nelson}},
  \bibnamefont{and} \bibinfo{author}{\bibfnamefont{C.}~\bibnamefont{Nisoli}},
  \bibinfo{journal}{New Journal of Physics} \textbf{\bibinfo{volume}{15}},
  \bibinfo{pages}{045009} (\bibinfo{year}{2013}).

\bibitem[{\citenamefont{Chern et~al.}(2013)\citenamefont{Chern, Morrison, and
  Nisoli}}]{Chern2013}
\bibinfo{author}{\bibfnamefont{G.-W.} \bibnamefont{Chern}},
  \bibinfo{author}{\bibfnamefont{M.~J.} \bibnamefont{Morrison}},
  \bibnamefont{and} \bibinfo{author}{\bibfnamefont{C.}~\bibnamefont{Nisoli}},
  \bibinfo{journal}{Phys. Rev. Lett.} \textbf{\bibinfo{volume}{111}},
  \bibinfo{pages}{177201} (\bibinfo{year}{2013}).

\bibitem[{\citenamefont{Mellado et~al.}(2012)\citenamefont{Mellado, Concha, and
  Mahadevan}}]{Mellado2012}
\bibinfo{author}{\bibfnamefont{P.}~\bibnamefont{Mellado}},
  \bibinfo{author}{\bibfnamefont{A.}~\bibnamefont{Concha}}, \bibnamefont{and}
  \bibinfo{author}{\bibfnamefont{L.}~\bibnamefont{Mahadevan}},
  \bibinfo{journal}{Physical review letters} \textbf{\bibinfo{volume}{109}},
  \bibinfo{pages}{257203} (\bibinfo{year}{2012}).

\bibitem[{\citenamefont{Han et~al.}(2008)\citenamefont{Han, Shokef, Alsayed,
  Yunker, Lubensky, and Yodh}}]{Han2008}
\bibinfo{author}{\bibfnamefont{Y.}~\bibnamefont{Han}},
  \bibinfo{author}{\bibfnamefont{Y.}~\bibnamefont{Shokef}},
  \bibinfo{author}{\bibfnamefont{A.~M.} \bibnamefont{Alsayed}},
  \bibinfo{author}{\bibfnamefont{P.}~\bibnamefont{Yunker}},
  \bibinfo{author}{\bibfnamefont{T.~C.} \bibnamefont{Lubensky}},
  \bibnamefont{and} \bibinfo{author}{\bibfnamefont{A.~G.} \bibnamefont{Yodh}},
  \bibinfo{journal}{Nature} \textbf{\bibinfo{volume}{456}},
  \bibinfo{pages}{898} (\bibinfo{year}{2008}).

\bibitem[{\citenamefont{M\"{o}ller and Moessner}(2006)}]{Moller2006}
\bibinfo{author}{\bibfnamefont{G.}~\bibnamefont{M\"{o}ller}} \bibnamefont{and}
  \bibinfo{author}{\bibfnamefont{R.}~\bibnamefont{Moessner}},
  \bibinfo{journal}{Phys. Rev. Lett.} \textbf{\bibinfo{volume}{96}},
  \bibinfo{pages}{237202} (\bibinfo{year}{2006}).

\bibitem[{\citenamefont{Chern and Mellado}(2013)}]{Chern2013magnetic}
\bibinfo{author}{\bibfnamefont{G.-W.} \bibnamefont{Chern}} \bibnamefont{and}
  \bibinfo{author}{\bibfnamefont{P.}~\bibnamefont{Mellado}},
  \bibinfo{journal}{arXiv preprint arXiv:1306.6154}  (\bibinfo{year}{2013}).

\bibitem[{\citenamefont{Castelnovo et~al.}(2008)\citenamefont{Castelnovo,
  Moessner, and Sondhi}}]{Castelnovo2008}
\bibinfo{author}{\bibfnamefont{C.}~\bibnamefont{Castelnovo}},
  \bibinfo{author}{\bibfnamefont{R.}~\bibnamefont{Moessner}}, \bibnamefont{and}
  \bibinfo{author}{\bibfnamefont{S.~L.} \bibnamefont{Sondhi}},
  \bibinfo{journal}{Nature} \textbf{\bibinfo{volume}{451}}, \bibinfo{pages}{42}
  (\bibinfo{year}{2008}).

\bibitem[{\citenamefont{Pauling}(1935)}]{Pauling1935}
\bibinfo{author}{\bibfnamefont{L.}~\bibnamefont{Pauling}},
  \bibinfo{journal}{Journal of the American Chemical Society}
  \textbf{\bibinfo{volume}{57}}, \bibinfo{pages}{2680} (\bibinfo{year}{1935}).

\bibitem[{\citenamefont{{Ramirez} et~al.}(1999)\citenamefont{{Ramirez},
  {Hayashi}, {Cava}, {Siddharthan}, and {Shastry}}}]{Ramirez1999}
\bibinfo{author}{\bibfnamefont{A.~P.} \bibnamefont{{Ramirez}}},
  \bibinfo{author}{\bibfnamefont{A.}~\bibnamefont{{Hayashi}}},
  \bibinfo{author}{\bibfnamefont{R.~J.} \bibnamefont{{Cava}}},
  \bibinfo{author}{\bibfnamefont{R.}~\bibnamefont{{Siddharthan}}},
  \bibnamefont{and} \bibinfo{author}{\bibfnamefont{B.~S.}
  \bibnamefont{{Shastry}}}, \bibinfo{journal}{Nature}
  \textbf{\bibinfo{volume}{399}}, \bibinfo{pages}{333} (\bibinfo{year}{1999}).

\bibitem[{\citenamefont{Gilbert et~al.}(2014)\citenamefont{Gilbert, Chern,
  Nisoli, and Schiffer}}]{Gilbert2014}
\bibinfo{author}{\bibfnamefont{I.}~\bibnamefont{Gilbert}},
  \bibinfo{author}{\bibfnamefont{G.-W.} \bibnamefont{Chern}},
  \bibinfo{author}{\bibfnamefont{C.}~\bibnamefont{Nisoli}}, \bibnamefont{and}
  \bibinfo{author}{\bibfnamefont{P.}~\bibnamefont{Schiffer}},
  \bibinfo{journal}{Submitted to Nature Phys.}  (\bibinfo{year}{2014}).

\end{thebibliography}

\end{document}